\documentclass[
    twocolumn,
    English,
    aps,
    pra,
    10pt,
    superscriptaddress,
    floatfix
]{revtex4-2}

\usepackage[T1]{fontenc}
\usepackage{newtxtext,newtxmath}
\usepackage{microtype}

\usepackage{graphicx}

\usepackage{booktabs}
\usepackage{multirow}

\usepackage[
    colorlinks=true,
    citecolor=blue,
    linkcolor=magenta
]{hyperref}

\hypersetup{
    pdfauthor={Anna Zhu},
    pdftitle={Understanding Intrinsic Loss in Thin-Film Lithium Niobate Ring Resonators via Adiabatic Coupling}
}

\begin{document}

	\title{Understanding Intrinsic Loss in Thin-Film Lithium Niobate Ring Resonators via Adiabatic Coupling}
	
	\author{Xinrui Zhu}\email{xzhu@g.harvard.edu}
    \affiliation{Harvard John A. Paulson School of Engineering and Applied Sciences, Harvard University, Cambridge, MA 02138, USA}

    \author{Hana K. Warner}
    \affiliation{Harvard John A. Paulson School of Engineering and Applied Sciences, Harvard University, Cambridge, MA 02138, USA}

    \author{Yunxiang Song}
    \affiliation{Harvard John A. Paulson School of Engineering and Applied Sciences, Harvard University, Cambridge, MA 02138, USA}
    \affiliation{Quantum Science and Engineering, Harvard University, Cambridge, MA 02138, USA}

    \author{Donald Witt}
    \affiliation{Harvard John A. Paulson School of Engineering and Applied Sciences, Harvard University, Cambridge, MA 02138, USA}

	\author{Marko Lončar}\email{loncar@g.harvard.edu}
	\affiliation{Harvard John A. Paulson School of Engineering and Applied Sciences, Harvard University, Cambridge, MA 02138, USA}

	\maketitle

\textbf{
Thin-film lithium niobate (TFLN) has emerged as a versatile integrated photonics platform, combining strong electro-optic and nonlinear effects. 
Among TFLN devices, ring resonators play a central role in filtering, modulation, and nonlinear optical processes. However, intrinsic loss -- which ultimately limits ring performance -- is most often summarized by single-valued metrics, and its statistical variability across resonances has received limited attention.
Here, we show that intrinsic loss rates in monolithic TFLN ring resonators follow a statistical distribution, comprising a baseline loss and a tail arising from discrete loss events.
This behavior is revealed by characterizing 2,233 resonances, using an adiabatic waveguide-ring coupling architecture that selectively excites the fundamental mode and yields clean spectra in the ultra–high-$Q_i$ regime.
We find the most probable intrinsic loss rate $\kappa_i = 2\pi \times 10.4$ MHz, indicating operation in a low-loss regime comparable to state-of-the-art thick silicon nitride platforms.
}

Thin-film lithium niobate (TFLN) has rapidly advanced 
as a leading platform for integrated photonics, 
    combining high electro-optic efficiency 
    with strong second-order nonlinearities
    \cite{zhu2021_integrated, boes2023_lithium,  hu2025_integrated}. 
This combination has enabled
broadband modulation \cite{wang2018_integrateda, xu2022_dualpolarization}, 
coherent signal transduction \cite{holzgrafe2020_cavity, warner2025_coherent}, 
frequency conversion \cite{mckenna2022_ultralowpowera, xin2025_wavelengthaccurate}, 
and chip-scale microcomb generation \cite{he2019_selfstarting, song2025_stable, nie2025_soliton, zhang2019_broadband}, 
    positioning TFLN as a key platform for classical and quantum photonic technologies \cite{song2025_integrated, hu2025_integrateda}. 
For many of these applications, 
ring resonators serve as a central building block \cite{zhang2017_monolithic, shams-ansari2022_reduceda, zhu2024_twentynine}, 
    where the enhancement of circulating optical fields enables 
    compact and efficient optical filtering, 
    modulation, and nonlinear interactions on chip.

Previous studies of TFLN ring resonators have explored different strategies 
to maximize intrinsic quality factor ($Q_i$).
For example, increasing the waveguide width has been shown to reduce optical mode overlap with sidewall roughness,
thereby suppressing scattering-related loss \cite{zhu2024_twentynine}.
Using this approach, $Q_i$ up to $\sim 2.9 \times 10^{7}$ 
have been demonstrated in wide-waveguide TFLN ring resonators.
However, previously used coupling schemes often either excite multiple mode families in the resonator 
or require local narrowing of the waveguide width at the coupling region \cite{zhu2024_twentynine, khalatpour_roughnesslimited}.
Here, we introduce an adiabatic coupling architecture
that enables selective excitation of the fundamental mode
while preserving a wide, low-scattering waveguide geometry throughout the entire ring \cite{ji2021_exploiting},
thereby enabling access to clean spectra in the ultra–high-$Q_i$ regime.

Analyzing a large ensemble of intrinsic loss data (2,233 resonances),
we observe that intrinsic loss exhibits a statistical distribution,
consisting of a Gaussian-like baseline and a tail associated with discrete loss events.
This perspective provides a complementary framework for understanding
resonance-to-resonance intrinsic loss variation in ring resonators,
by explicitly characterizing the statistical structure of the loss distribution.
Furthermore, the most probable intrinsic loss rate extracted from this ensemble,
$\kappa_i = 2\pi \times 10.4~\mathrm{MHz}$ (corresponding to $Q_i \sim 1.9 \times 10^{7}$),
indicates that thin-film lithium niobate operates in a low-loss regime comparable to
state-of-the-art thick silicon nitride platforms.

\begin{figure*}[htbp]
    \centering
    \includegraphics[page=1,width=90mm]{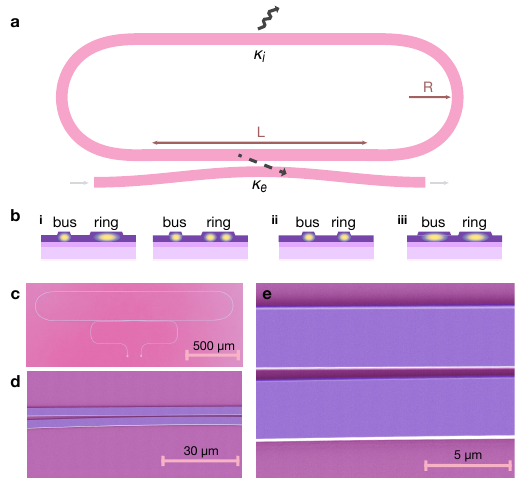}
    \caption{\textbf{Mode-selective coupling in thin-film lithium niobate (TFLN) ring resonators.}  
    \textbf{(a)} Schematic of a bus-coupled racetrack resonator. 
    The straight-section length ($L$) and bend radius ($R$) define the resonator geometry. 
    The intrinsic loss and external coupling rates are denoted by $\kappa_i$ and $\kappa_e$, respectively.
    \textbf{(b)} Three representative coupling strategies. 
    In conventional approaches, light from a single-mode bus waveguide is either 
    \textbf{(i)} coupled into a wide, multimode ring, thereby exciting multiple modes, or 
    \textbf{(ii)} the ring is tapered to a single-mode width in the coupling region, which can introduce additional scattering loss. 
    In this work (\textbf{iii}), we use a wide, low-loss bus waveguide and ring resonator, but introduce an adiabatic coupler that gradually bends the bus waveguide toward the ring, enabling excitation of only the fundamental mode while preserving the wide ring geometry.
    \textbf{(c)} Optical microscope image of a fabricated racetrack resonator incorporating an adiabatic coupling region for selective fundamental-mode excitation.
    \textbf{(d)} False-colored scanning electron microscope (SEM) image of the coupling region.
    \textbf{(e)} Zoomed-in SEM image of the coupling gap.
    }
    \label{Fig_1}
\end{figure*}
 
\subsection*{Ring Resonators and Loss Mechanisms}

In this work, we employ a racetrack resonator geometry,
a common variant of ring resonators with straight sections, as shown in Fig.~\ref{Fig_1}a.
By forming a closed waveguide loop, 
optical fields that satisfy the resonance condition
can be coherently stored inside the cavity,
thereby enhancing light--matter interactions on chip \cite{yariv2002_critical, bogaerts2012_silicon}. The total energy decay rate of the cavity is typically given by
\begin{equation}
\kappa = \kappa_i + \kappa_e,
\label{eq:kappa_sum}
\end{equation}
where $\kappa_i$ denotes the intrinsic loss rate
associated with propagation inside the ring,
and $\kappa_e$ denotes the coupling rate
between the ring and the bus waveguide.
The corresponding quality factors are defined as
\begin{equation}
Q = \frac{\omega_0}{\kappa}, \qquad
Q_i = \frac{\omega_0}{\kappa_i}, \qquad
Q_e = \frac{\omega_0}{\kappa_e},
\label{eq:Q_kappa}
\end{equation}
where $\omega_0$ is the resonance angular frequency. These provide a convenient, dimensionless parametrization of the cavity loss channels.
In our discussion, $Q_i$ and $\kappa_i$ are used interchangeably to describe the same physical loss process, while all statistical analyses are performed in terms of $\kappa_i$, which is independent of the resonance frequency.

To identify the origins of intrinsic propagation loss, 
we consider two dominant loss mechanisms. 
The first mechanism is absorption loss, 
    in which optical energy is irreversibly dissipated as heat within the material. 
In this work, we consider only linear material absorption under low-power operation, 
where nonlinear optical loss mechanisms in TFLN are negligible \cite{bache2012_review,xu2021_mitigating}.
The absorption strength is determined by 
the imaginary part of the refractive index of the constituent layers 
(e.g., the LN device layer and the underlying \(\mathrm{SiO_2}\) layer), 
    and may be mitigated through material and structural engineering.

The second mechanism is scattering loss, 
    in which optical energy is coupled into radiation modes or unwanted higher-order modes 
    and is effectively lost. 
This scattering typically arises from nanoscale roughness 
    at etched sidewalls and waveguide surfaces, 
    as well as from other perturbations to the optical mode. 
It is strongly dependent on the mode field distribution 
    and therefore can be engineered through 
    careful waveguide geometry design and fabrication optimization.
In particular, the coupling region is an important site for such scattering pathways.
As light transfers between the bus waveguide and the ring, 
geometric or refractive-index perturbations can cause mode mismatch, 
introducing an additional parasitic loss channel that contributes to the intrinsic loss rate \(\kappa_i\) 
rather than the extrinsic coupling rate \(\kappa_e\).

A common strategy to suppress scattering loss in TFLN ring resonators 
is to employ a wide waveguide for the ring, 
where the fundamental mode is positioned away from the etched sidewalls. 
This design minimizes the mode overlap with surface roughness 
and enables ultra-high intrinsic quality factors (\(Q_i\)). 
However, direct coupling from a single-mode bus waveguide into a wide ring waveguide (Fig.~\ref{Fig_1}b(i)) can excite multiple mode families
\cite{zhu2024_twentynine}.
A workaround is to locally narrow both the bus and the ring 
waveguides in the coupling region to enforce single-mode coupling (Fig.~\ref{Fig_1}b(ii)). 
This narrowing increases the mode overlap with the sidewalls, 
which can, in principle, reintroduce scattering loss and limit the achievable \(Q_i\)  \cite{khalatpour_roughnesslimited,kordts2016_higher}.

In this work, we preserve the wide-waveguide geometry of the ring 
to maintain its low-scattering advantage (Fig.~\ref{Fig_1}b(iii)) \cite{ji2021_exploiting}. 
By introducing an adiabatic bend, 
the bus waveguide remains effectively single-mode in the coupling region, 
    selectively exciting only the fundamental mode of the ring.
This design enables clean single-mode resonance spectra and ultra-high intrinsic quality factors, 
    uniting the benefits of both low scattering and mode selectivity.

\begin{figure*}[htbp]
    \centering
    \includegraphics[page=1,width=90mm]{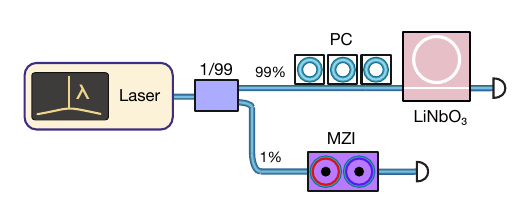}
    \caption{\textbf{Measurement setup.}
    A tunable laser is split using a 1/99 splitter, where the 99\% arm is sent through a polarization controller (PC) to the device under test, and the 1\% arm is routed to a Mach--Zehnder interferometer (MZI) to generate calibration fringes for wavelength referencing.}
    \label{Fig_2}
\end{figure*}

\subsection*{Device Design and Characterization}

Our devices employ a racetrack resonator geometry with Euler bends 
of radius \(120~\mu\mathrm{m}\), 
as illustrated in Fig.~\ref{Fig_1}c. 
The waveguide width is \(3~\mu\mathrm{m}\), 
and the straight section length is \(1.4 ~\mathrm{mm}\),
corresponding to a free spectral range (FSR) of approximately \(33~\mathrm{GHz}\).
Compared with our previous ultra-large rings with an FSR of around 6~GHz (\(4.5~\mu\mathrm{m}\) waveguide width, \(1~\mathrm{cm}\) straight-section length, \(200~\mu\mathrm{m}\) bending radius, and maximum \(Q_i\) of \(2.9\times10^7\)) \cite{zhu2024_twentynine},
this compact geometry provides a practical balance between propagation loss and overall device footprint. 

The devices were fabricated on chips cleaved from 4-inch X-cut TFLN on insulator wafers purchased from NanoLN.
The wafer stack consists of a 600~nm device layer bonded to a 4.7~$\mu$m thermal SiO$_2$ layer on a silicon substrate.
Waveguide patterns were defined by electron-beam lithography using hydrogen silsesquioxane (HSQ) resist and subsequently transferred into the LN layer by Ar plasma etching.
Post-etch redeposition was removed through a heated SC-1 clean, followed by diluted HF and piranha treatments to ensure surface cleanliness.
Finally, the devices were annealed in oxygen at $520~^\circ\mathrm{C}$ for two hours to mitigate plasma-induced damage and reduce optical absorption.
This straightforward monolithic process provides a robust and reproducible fabrication approach that has been broadly applied in our TFLN photonic device developments.
As shown in Fig.~\ref{Fig_1}d--e, 
SEM micrographs reveal well-defined sidewalls 
and coupling regions, 
highlighting the precision and reliability of our fabrication process.

Optical characterization was performed using the setup schematically shown in Fig.~\ref{Fig_2}.
Light from a tunable continuous-wave (CW) laser was split:
99\% of the power was directed to the TFLN resonator under test,
and 1\% was routed to a reference Mach–Zehnder interferometer (MZI).
The transmitted light from the resonator 
    was detected by a high-speed photodetector,
while the MZI output provided periodic interference fringes 
    that served as a frequency reference to correct laser sweep nonlinearity.
The detector dark voltage, measured separately, was subtracted to correct baseline offsets.
With this calibration procedure, 
transmission spectra were obtained 
across the 1480–1600~nm wavelength range for all resonators.
These calibrated spectra serve as the basis for extracting
the intrinsic quality factors used in the following analysis.

\begin{figure*}[htbp]
    \centering
    \includegraphics[page=1,width=180mm]{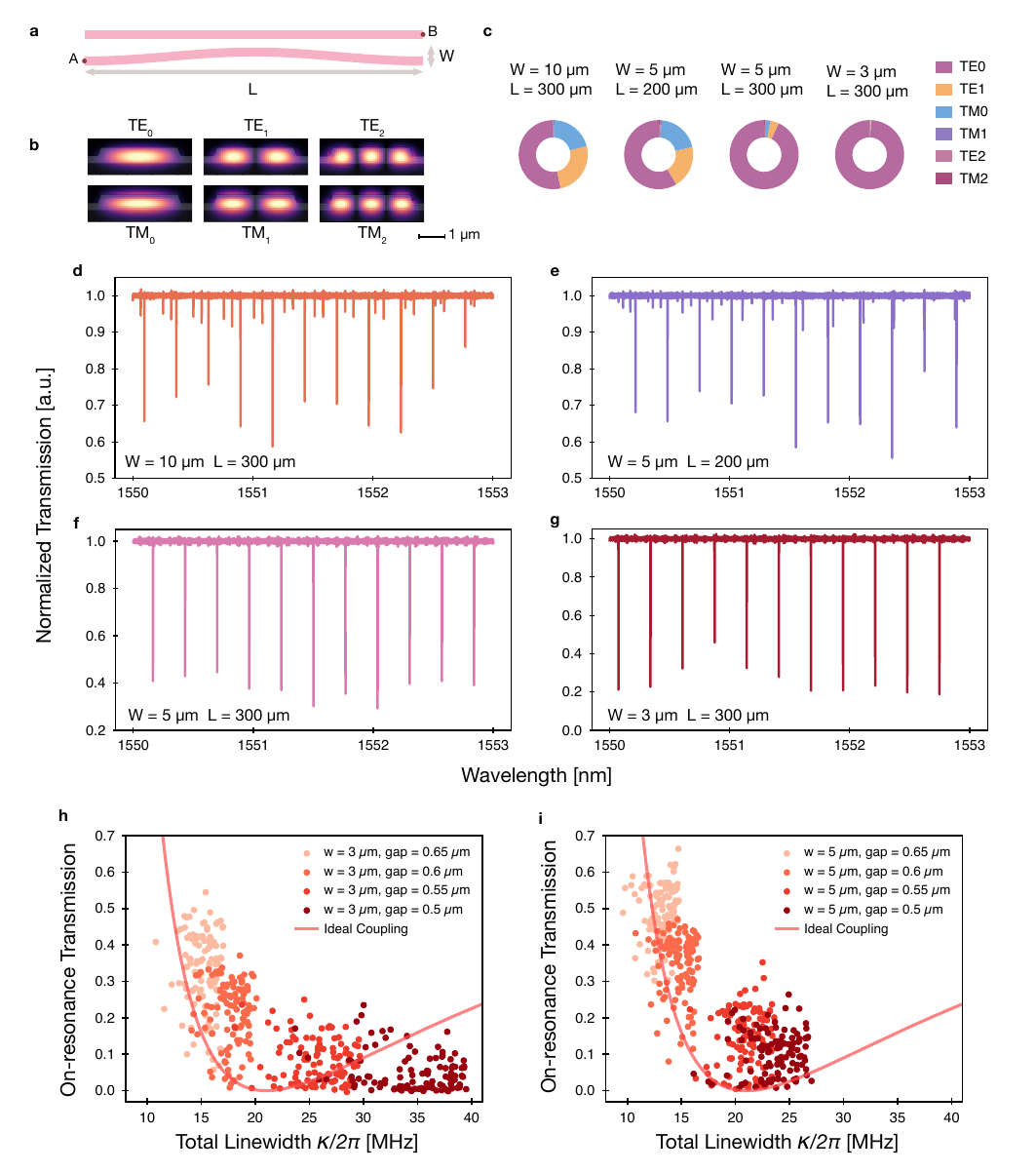}
    	\caption{\textbf{Mode-selective coupling enabled by an adiabatic coupling geometry.}
            \textbf{(a)} Schematic of the coupling region.
            The parameter $w$ denotes the lateral offset of the bus-waveguide center between the entrance and the midpoint of the coupling region, and $L$ is the total coupler length.
            \textbf{(b)} Simulated guided modes supported in a $3~\mu\mathrm{m}$-wide TFLN waveguide (TE$_0$–TE$_2$, TM$_0$–TM$_2$).
            \textbf{(c)} Simulated modal power fractions in the ring waveguide for coupling geometries with different $(w, L)$. 
            The fundamental mode is excited at the bus-waveguide entrance (point A in panel (a)), while the mode composition is evaluated at the exit of the ring-waveguide coupling region (point B in panel (a)).
            \textbf{(d--g)} Normalized transmission spectra measured for the corresponding fabricated devices.
            \textbf{(h--i)} On-resonance transmission versus total linewidth $\kappa/2\pi$ for measured resonances.
            The solid curve indicates the theoretical linewidth–extinction relation for ideal coupling, assuming an intrinsic loss rate of $\kappa_i/2\pi = 10.5~\mathrm{MHz}$.
            Scattered points represent experimental data acquired for different coupling gaps, in the wavelength range of 1520-1580~nm.
        	}
    \label{Fig_3}
\end{figure*}

\subsection*{Mode Selection in Adiabatic Coupling}

The guiding principle of our coupler design is \textit{adiabaticity}, in which a system evolves gradually enough to remain in its instantaneous eigenstate without inducing unwanted transitions.
In the context of integrated photonics, an adiabatic coupler ensures that the optical mode evolves smoothly along the propagation direction, preserving its mode identity and minimizing reflection, scattering, and unintended mode conversion.

The coupler geometry is illustrated in Fig.~\ref{Fig_3}a. 
The bus waveguide follows a squared-sinusoidal trajectory, 
providing a smooth and symmetric variation of the coupling gap 
with zero slope at both ends to satisfy the adiabatic condition at the boundaries; 
no claim of global optimality is made for this specific trajectory choice.
Before and after the coupling region, the bus waveguide is tapered 
from a single-mode width of \(0.8~\mu\mathrm{m}\) to \(3~\mu\mathrm{m}\).
Importantly, within the coupling region, both the bus and the ring waveguides have the same \(3~\mu\mathrm{m}\) width,
ensuring phase-matched coupling between their fundamental modes while maintaining a gradual modal transition.
This configuration is designed to provide minimal perturbation and broad coupling bandwidth.

To investigate the mode behavior in the coupling region, 
we combined eigenmode analysis with three-dimensional 
finite-difference time-domain (FDTD) simulations using Tidy3D.
As shown in Fig.~\ref{Fig_3}b, 
a \(3~\mu\mathrm{m}\)-wide waveguide supports multiple 
TE$_0$–TE$_2$ and TM$_0$–TM$_2$ modes.
In our experiment, light is injected through a grating coupler designed for TE\(_0\) excitation 
near \(1550~\mathrm{nm}\), and the optical mode entering the coupling region is therefore primarily TE\(_0\).
In the simulation, we placed the mode source at the entrance (left side of the bus waveguide) 
and monitored the mode composition at the exit (right side of the ring waveguide) 
for different coupling geometries.
Mode purity is quantified by the simulated power fraction in  TE\(_0\) at the ring-waveguide exit.
As shown in Fig.~\ref{Fig_3}c,
shorter coupling lengths (\(L = 200~\mu\mathrm{m}\)) or excessive lateral offsets (\(w = 10~\mu\mathrm{m}\)) 
lead to substantial excitation of higher-order modes,
indicating a breakdown of adiabatic mode evolution.
In contrast, for longer coupling lengths (\(L = 300~\mu\mathrm{m}\)) 
and moderate lateral offsets (\(w = 3~\mu\mathrm{m}\)–\(5~\mu\mathrm{m}\)),
the simulated mode profiles remain nearly pure.
We note that, in principle, adiabatic mode evolution could also be achieved with larger lateral separations at longer coupling lengths; 
in this work, however, we focus on compact coupling geometries and do not discuss these extended designs further.

The measurements corroborate the simulation results.
Fig.~\ref{Fig_3}d--g present the transmission spectra of devices 
with different coupling geometries.
Devices with shorter or more abrupt coupling sections
exhibit multiple sets of resonances, reflecting the presence of higher-order mode families.
By contrast, devices employing longer and more adiabatic coupling regions
show a single, well-defined resonance series with clearly isolated dips,
demonstrating that these geometries effectively suppress higher-order 
mode excitation in the ring.
A quantitative evaluation supporting the mode purity discussion is detailed in the Supplementary Information.

Coupling ideality is a figure of merit that quantifies the fraction of optical power transferred into the desired mode rather than being coupled into undesired modes or lost through parasitic channels \cite{spillane2003_ideality, pfeiffer2017_coupling, wang2024_coupling}.
Notably, single-mode spectra do not necessarily imply ideal coupling.
A detailed discussion of the coupling ideality framework is provided in the Supplementary Information.
In the ideal coupling regime, parasitic loss is negligible, and the relationship between the total resonance linewidth and extinction depth follows the theoretical transmission model, as shown by the solid curves in Fig.~\ref{Fig_3}h--i.
This behavior exhibits a continuous transition from the under-coupled regime through critical coupling to the over-coupled regime as the coupling strength increases \cite{pfeiffer2017_coupling}.

We examine the linewidth–extinction relationship for geometries with $L = 300~\mu\mathrm{m}$ and $w = 3\text{–}5~\mu\mathrm{m}$, as shown by the scattered data points in Fig.~\ref{Fig_3}h--i.
Different coupling gaps probe a range of coupling strengths, and only resonances with $\kappa$ below the median value are included to isolate intrinsic loss variations.
We observe that devices with $w = 3~\mu\mathrm{m}$ exhibit deviations from the ideal transmission behavior as the coupling strength increases, particularly for a gap of $0.5~\mu\mathrm{m}$.
The smaller separation between the bus and ring across the entire coupling region increases sensitivity to fabrication-induced nonuniformities and enhances the integrated mode overlap with parasitic channels.
We therefore adopt a coupler geometry with $w = 5~\mu\mathrm{m}$ for subsequent high-$Q_i$ studies, as it operates farther from the regime where coupling non-ideality begins to dominate.

\subsection*{Statistical Distributions of Intrinsic Loss}

\begin{figure*}[htbp]
    \centering
    \includegraphics[page=1,width=180mm]{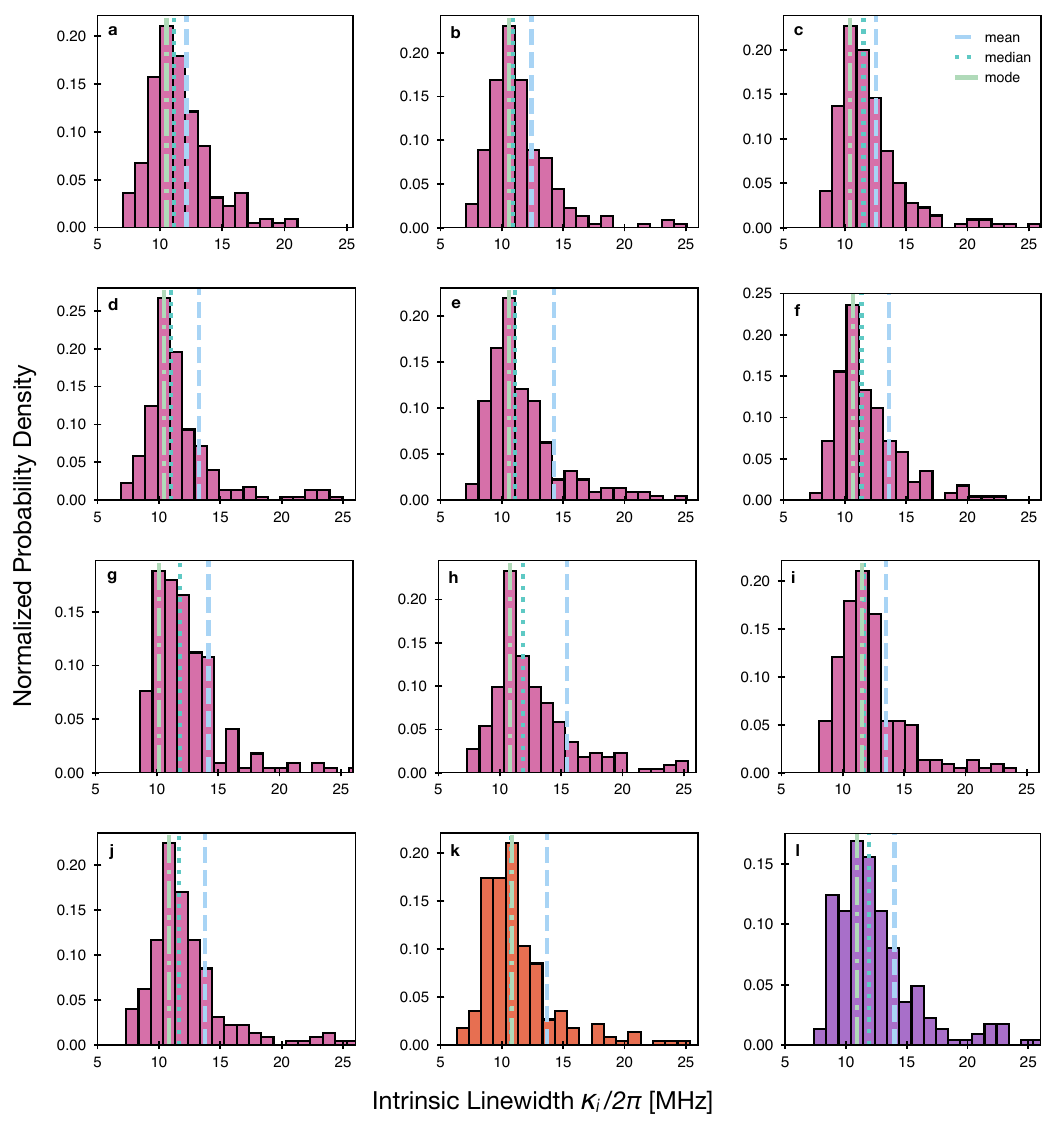}
    	\caption{\textbf{Statistical distribution of intrinsic loss rates \(\kappa_{i}\) across multiple resonators.}
        \textbf{(a--l)} Normalized histograms of the intrinsic linewidth \(\kappa_{i}/2\pi\) for twelve ring resonators fabricated with the same coupling design \((w = 5~\mu\mathrm{m}, L = 300~\mu\mathrm{m})\).
        Resonances within the 1520–1580~nm wavelength range are extracted and fitted to obtain \(\kappa_{i}\).
        Vertical guides indicate the mean (light blue, long-dashed), median (teal, dotted), and mode (light green, dash–dot) of each device's distribution.
        The x-axis range is fixed from 5 to 26~MHz for consistent comparison.
        Devices (a--j) are fabricated on the same chip, while (k) and (l) are fabricated on two additional chips. The systematic separation between mean, median, and mode across all devices highlights the intrinsically asymmetric nature of loss statistics.
	}
    \label{Fig_4}
\end{figure*}

\begin{figure*}[htbp]
    \centering
    \includegraphics[page=1,width=175mm]{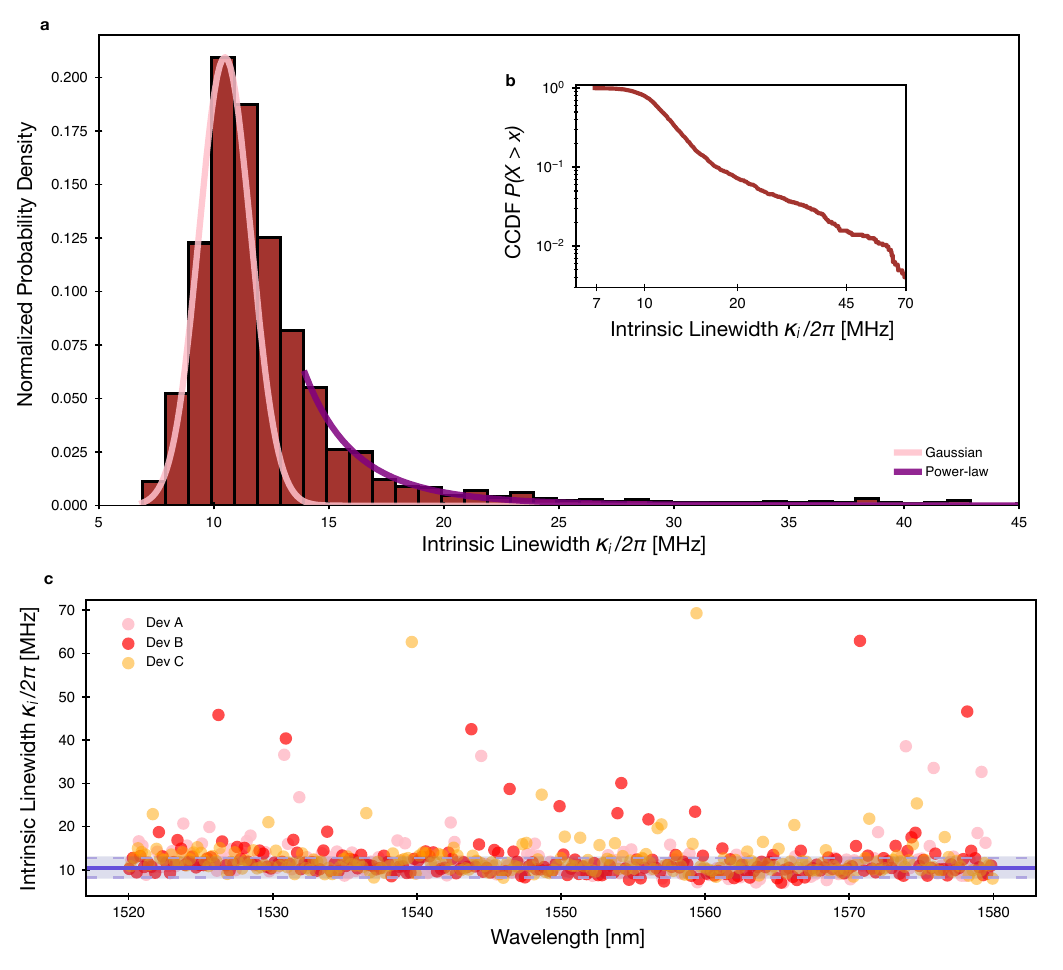}
    \caption{\textbf{Statistical characterization and modeling of intrinsic loss.}
    \textbf{(a)} Combined normalized histogram of intrinsic linewidth 
    $\kappa_{i}/2\pi$ from ten devices (2,233 resonances). 
    The distribution exhibits a Gaussian-like core (pink) together with a 
    truncated power-law-like tail (purple). 
    \textbf{(b)} Complementary cumulative distribution function (CCDF) of the 
    same dataset, showing a deviation from Gaussian behavior in the 
    high-loss regime, with approximately power-law-like decay over an 
    intermediate range followed by a gradual cutoff at large linewidths. 
    \textbf{(c)} Intrinsic linewidth $\kappa_{i}/2\pi$ versus wavelength for 
    three representative devices (corresponding to devices in Fig.~4(a--c)). 
    Each point corresponds to a single resonance. 
    The shaded band denotes the Gaussian core ($\mu \pm 2\sigma$) extracted 
    from the global fit in (a), illustrating that most resonances cluster 
    around the baseline loss, while high-loss events occur sporadically 
    across the wavelength range.
    For clarity, the histogram is displayed up to 
    $\kappa_i/2\pi = 45~\mathrm{MHz}$, and the CCDF is displayed up to 
    $\kappa_i/2\pi = 70~\mathrm{MHz}$, focusing on the regime with robust 
    statistical support.
    }
    \label{Fig_5}
\end{figure*}

To quantify the intrinsic loss rate of our racetrack resonators,
we performed a statistical analysis across twelve nominally identical 
devices sharing the same ring geometry (\(3~\mu\mathrm{m}\) waveguide width, 
\(33~\mathrm{GHz}\) FSR) and optimized coupling design
(\(w = 5~\mu\mathrm{m}\), \(L = 300~\mu\mathrm{m}\)), as summarized in 
Fig.~\ref{Fig_4}.
Devices~(a–j) were patterned and processed on the same chip, whereas
devices~(k) and~(l) were fabricated on two additional chips from 
independent runs, providing a useful check on chip-to-chip 
reproducibility.
To ensure reliable distinction between $\kappa_i$ and $\kappa_e$,
we include only devices operating 
in the moderately undercoupled regime
(gaps of $\sim 0.60\text{–}0.65~\mu\mathrm{m}$),
where the fits are robust and directly comparable.
The complete fitting formalism and representative examples are provided in the Supplementary Information.
This consistently extracted set of intrinsic loss rates
provides a solid basis for examining the statistical behavior
of intrinsic loss in TFLN ring resonators.

With this dataset in place, we examine the statistical distribution
of intrinsic loss rates \(\kappa_i\) across the twelve devices.
Because the intrinsic loss exhibits a non-symmetric distribution,
single-valued summary statistics are insufficient to characterize device performance.
For each device, we extract the mean, median, and mode of the
\(\kappa_i\) distribution, as these complementary statistics capture
different aspects of the underlying loss behavior.
The \textit{mean} intrinsic loss rate reflects the arithmetic average,
but is strongly influenced by probabilistic, high-loss resonances with
exceptionally large \(\kappa_i\).
The \textit{median}, by contrast, is far less sensitive to outliers
and provides a more robust indicator of the central behavior of 
intrinsic loss across a device.
The \textit{mode}, defined as the most probable intrinsic loss rate,
captures the value most frequently realized across the resonance 
ensemble.
In this work, the mode is taken as the intrinsic loss rate associated with the highest-probability bin of the empirical histogram, using a fixed bin width applied consistently across all devices. Our conclusions are robust under reasonable variations of histogram bin width. 

Although each device-level histogram reflects the sensitivity of
\(\kappa_i\) to local fabrication variations, aggregating all 2,233
resonances from ten devices (Fig.~\ref{Fig_4}a--j) from the same chip yields a much smoother distribution that reveals the underlying statistical structure, as shown in Fig.~\ref{Fig_5}a.
For the aggregated dataset, the mean \(\kappa_i/2\pi\) is 
\(13.5~\mathrm{MHz}\), the median is \(11.4~\mathrm{MHz}\), 
and the mode is \(10.4~\mathrm{MHz}\).
The minimum observed \(\kappa_i/2\pi\) is approximately \(7~\mathrm{MHz}\).

The aggregated linewidth distribution exhibits a Gaussian-like main lobe 
together with a truncated power-law tail component 
(Fig.~\ref{Fig_5}a), 
suggesting the presence of both baseline loss and additional high-loss events.
The statistical fitting pipeline and robustness validation are presented in the Supplementary Information.

The Gaussian component reflects the intrinsic baseline loss, arising from many small, uncorrelated contributions
    such as surface roughness, 
    material absorption, 
    and other minor fabrication- or material-related imperfections.
The Gaussian core is centered at $\mu \approx 10.48$~MHz, in close agreement with the most-probable value extracted from the histogram.
Since all loss channels contribute positively, 
any additional mechanism simply adds to the baseline loss 
and shifts the distribution toward larger linewidths.

The tail portion, by contrast, corresponds to occasional high-loss events. 
Such discrete events may arise when several factors — including the optical field distribution, local defect configurations, and the possible presence of higher-order or radiation modes at certain wavelengths — coincide in a manner that enhances loss.
In realistic devices, the optical fields exhibit frequency-dependent variations both in their transverse profiles and along the ring. 
Longitudinal spatial variations, in particular, may arise from sidewall roughness and other imperfections that induce backscattering, leading to partially standing-wave-like resonant modes.
Consistent with this qualitative interpretation, the complementary cumulative distribution function (CCDF), which directly quantifies the likelihood of observing resonances with intrinsic loss rates exceeding a given threshold, exhibits a power-law-like decay over an intermediate range, followed by a gradual cutoff at large linewidths, as shown in Fig.~\ref{Fig_5}b.

While the histogram and CCDF summarize the statistics, 
Fig.~\ref{Fig_5}c shows how these mechanisms manifest at the device level. 
Across three representative resonators, most linewidths lie close to the 
Gaussian core (shaded region), whereas the high-loss events 
occur sporadically from resonance to resonance across the wavelength range, 
consistent with their origin in 
inherently probabilistic, mode-dependent interactions.
This device-level behavior points to a distinction between the factors that 
set the most probable \,\(Q_i\) and those that determine the average \,\(Q_i\).  
Because the latter is shaped strongly by tail events, its improvement is 
more closely tied to mitigating rare but systematic, conditionally activated loss channels than to 
further refining the baseline loss 
— an observation that offers a more general perspective on 
loss-engineering strategies.

\subsection*{Conclusion and Outlook}

We demonstrate ultra--high--$Q_i$ thin-film lithium niobate (TFLN) ring resonators 
enabled by an adiabatic coupling architecture that enforces fundamental-mode operation.
This design produces clean spectra with robust suppression of higher-order modes, 
allowing intrinsic loss to be extracted with high confidence across a broad ensemble of resonances.
From twelve devices spanning three fabrication runs, 
we report the mean, median, and most probable intrinsic loss with full statistical transparency for monolithic TFLN.
Building on these device-level results, our dataset across 2,233 resonances enables a distribution-level description of intrinsic loss in TFLN ring resonators:
a Gaussian baseline shaped by numerous small, uncorrelated loss components,
overlaid with discrete tail events associated with probabilistically activated loss channels.

In addition to recasting loss engineering into a unified physical–statistical perspective, our results place TFLN alongside the leading low-loss integrated photonic platforms.
In thick silicon nitride, state-of-the-art intrinsic loss is realized through 
either Damascene reflow or surface–mechanical polishing, yielding 
most-probable $Q_i$ values of $3\times10^7\,\text{--}\,4\times10^7$ \cite{liu2021_highyield, ji2017_ultralowloss}.
By comparison, our most-probable $Q_i$ values of $\approx 2\times10^7$ are achieved through a straightforward fabrication process.

Looking ahead, the combination of high–$Q_i$ single-mode operation with electro-optic and all-optical nonlinearities opens a broader horizon for integrated photonics in TFLN.


\section*{Funding}

National Science Foundation (NSF) (EEC-1941583, ECCS-2407727);
Defense Advanced Research Projects Agency (DARPA) (HR0011-24-2-0360);
Air Force Office of Scientific Research (AFOSR) (FA955024PB004);
Naval Air Warfare Center Aircraft Division (N6833522C0413);
Air Force Lifecycle Management Center (FA8702-15-D-0001);
National Research Foundation of Korea (NRF) (NRF-2022M3K4A1094782);
Amazon Web Services (AWS) (A50791, A60290).

\section*{Author contributions}

X.Z. conceived the project, developed the device concept and experimental plan, performed the design, fabrication, measurements, and data analysis, and prepared the manuscript with input from all authors. 
H.W. assisted with device fabrication and contributed to the development of the resonance–spectrum fitting framework. 
Y.S. assisted with the optical measurements and led the development of the MZI-based wavelength calibration code. 
D.W. assisted with the HSQ dose optimization experiments. 
M.L. supervised the project and provided guidance on the research direction and the manuscript.

\section*{Acknowledgments}
Device fabrication was carried out at the Harvard University Center for Nanoscale Systems (CNS).

\section*{Competing interests}
M.L. is a co-founder of a company that works on thin-film lithium niobate. The authors declare no other competing interests.

\section*{Data availability} 
Data underlying the results presented in this paper are available upon reasonable request.

\clearpage

\appendix

\section*{Supplementary Information} 

\section{Resonance Fitting Model}

We fit the measured transmission intensity using
\begin{equation}
I(\omega)=|S(\omega)|^2,\qquad 
S(\omega)=A(\omega)+B\,t(\omega),
\label{eq:S_def}
\end{equation}
where $A(\omega)$ is a slowly varying complex background term (modeled as a linear function of frequency), $B$ is a complex-valued coefficient that scales the cavity transmission amplitude $t(\omega)$.
The coherent interference between the background and cavity response reproduces the observed Fano-like asymmetry.

For non-split resonances,
\begin{equation}
t(\omega)=1-\frac{\kappa_e}{\frac{\kappa}{2}+i(\omega-\omega_0)},
\label{eq:t_nonsplit}
\end{equation}
where $\kappa=\kappa_i+\kappa_e$ is the total field decay rate, $\omega$ denotes the angular frequency of the input field, and $\omega_0$ is the resonance angular frequency of the cavity mode under consideration.

For split resonances induced by backscattering between clockwise and counterclockwise modes,
we use
\begin{equation}
t(\omega)=1-\frac{\kappa_e\left(\frac{\kappa}{2}+i(\omega-\omega_0)\right)}
{\left(\frac{\kappa}{2}+i(\omega-\omega_0)\right)^2+\left(\frac{\kappa_c}{2}\right)^2},
\label{eq:t_split}
\end{equation}
where $\kappa_c$ denotes the coherent backscattering-induced coupling rate between the clockwise and counterclockwise
modes, responsible for mode splitting \cite{li2013_unified}. 

Parameters are extracted by fitting the transmission spectra as a function of frequency $f$ using nonlinear least-squares optimization
(Levenberg-Marquardt algorithm) implemented via the \texttt{lmfit} Python package. We convert
between $f$ and $\omega$ using $\omega=2\pi f$.
Accordingly, linewidths extracted in Hz correspond to $\kappa/2\pi$.

\begin{figure*}[htbp]
    \centering
    \includegraphics[page=1,width=175 mm]{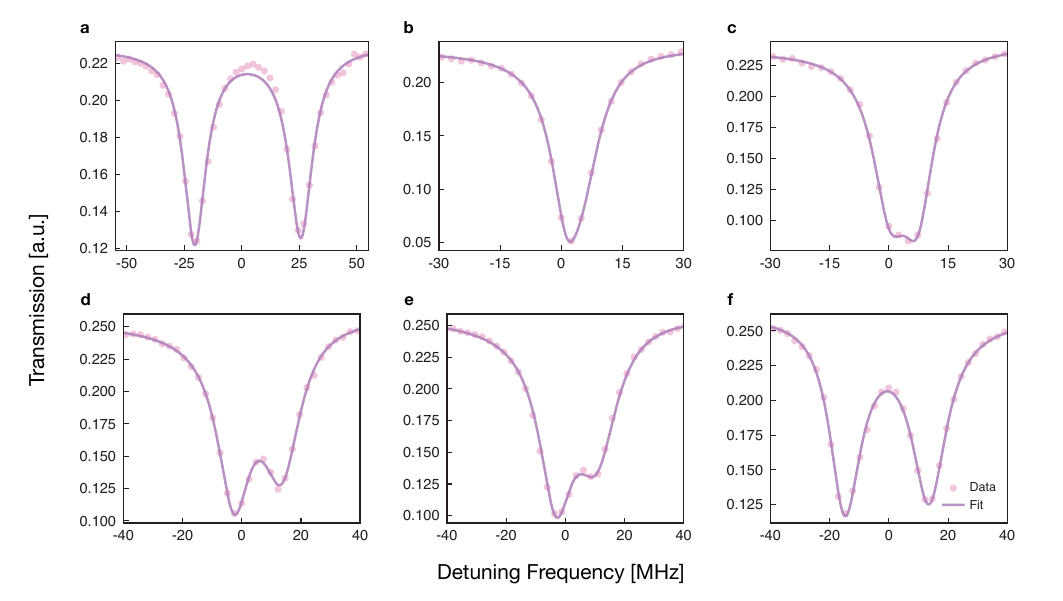}
    	\caption{\textbf{Representative resonance spectrum fitting (a–f).}
    Measured transmission spectra (circles) and corresponding fits (solid lines) for representative resonances. 
    The extracted fitting parameters are summarized in Table \ref{Table_S1}.}
    \label{Fig_S1}
\end{figure*}

All resonances are fitted using the split-mode model described above. 
In the absence of visible splitting, the model reduces to the single-mode limit as $\kappa_c \to 0$. 
Representative fitting examples are shown in Fig. \ref{Fig_S1}, demonstrating the model performance across different effective coupling strengths.

\begin{table}[htbp]
\centering
\caption{\bf Extracted fitting parameters for representative resonances shown in Fig. \ref{Fig_S1}. 
Linewidths are reported as $\kappa/2\pi$ in MHz.}
\begin{tabular}{ccccccc}
\hline
Resonance & $\lambda$ (nm) & $\kappa/2\pi$ & 
$\kappa_i/2\pi$  & $\kappa_e/2\pi$ & 
$\lvert \kappa_c \rvert/2\pi$ & $Q_i$ \\
\hline
(a) & 1565.15 &  12.0 & 8.9 &  3.1 &  46.0 & $2.15\times10^{7}$ \\
(b) & 1565.43  &  10.3 &  7.1 & 3.2 &  4.4 & $2.70\times10^{7}$ \\
(c) & 1565.70  &  10.3 &  7.1 & 3.2 &  7.5 & $2.70\times10^{7}$ \\
(d) & 1565.97  &  14.7 & 10.7 & 4.0 & 16.7 & $1.78\times10^{7}$ \\
(e) & 1566.24 &  15.6 & 11.3 & 4.3 & 14.5 & $1.70\times10^{7}$ \\
(f) & 1566.52 &  13.5 & 9.6 & 4.0 & 28.8 &  $2.00\times10^{7}$ \\
\hline
\end{tabular}
\label{Table_S1}
\end{table}

\section{Quantitative Analysis of Mode Purity}

In the main text, qualitative evidence of single-family behavior is provided through representative simulations and measured spectra. Here, we introduce a quantitative framework to evaluate mode purity using clearly defined metrics from both simulation and experimental perspectives.

In the simulation-based evaluation, mode purity is quantified by calculating the fractional power carried by each supported eigenmode at the output cross-section of the coupling region. The modal fractions are normalized to the total transmitted power at the monitor plane, thereby isolating modal selectivity from overall coupling efficiency. The relative modal fractions are directly compared across different coupling gaps and wavelengths.

While the absolute coupling efficiency varies with gap and wavelength, the dominance of the selected mode remains consistent.
The simulation data for the four geometries are summarized in Tables \ref{Table_S2}-\ref{Table_S5}. Modal fractions are defined as $f_i = P_i / \sum_j P_j$, where $P_j$ includes TE$_0$, TE$_1$, TM$_0$, TM$_1$, TE$_2$, and TM$_2$. Fig. \ref{Fig_3}c in the main text corresponds to the simulation results for a 0.6~$\mu$m gap at 1550~nm for these geometries. 

\begin{table}[h]
\centering
\caption{Simulated modal fractions at the output monitor for $w = 3~\mu$m and $L = 300~\mu$m.}
\begin{tabular}{c c c c c}
\hline
Gap ($\mu$m) & Wavelength (nm) & TE0 (\%) & TE1 (\%) & TM0 (\%) \\
\hline

\multirow{3}*{0.5} 
 & 1530 & 98.96 & 0.42 & 0.60 \\
 & 1550 & 99.40 & 0.42 & 0.14 \\
 & 1570 & 99.56 & 0.25 & 0.15 \\

\hline

\multirow{3}*{0.6} 
 & 1530 & 99.11 & 0.34 & 0.51 \\
 & 1550 & 99.55 & 0.33 & 0.11 \\
 & 1570 & 99.70 & 0.19 & 0.11 \\

\hline

\multirow{3}*{0.7} 
 & 1530 & 98.98 & 0.38 & 0.61 \\
 & 1550 & 99.49 & 0.33 & 0.13 \\
 & 1570 & 99.63 & 0.21 & 0.13 \\

\hline
\end{tabular}
\label{Table_S2}
\end{table}
\begin{table}[h]
\centering
\caption{Simulated modal fractions at the output monitor for $w = 5~\mu$m and $L = 300~\mu$m.}
\begin{tabular}{c c c c c}
\hline
Gap ($\mu$m) & Wavelength (nm) & TE0 (\%) & TE1 (\%) & TM0 (\%) \\
\hline

0.5 & 1530 & 91.34 & 3.24 & 5.34 \\
    & 1550 & 93.79 & 4.19 & 1.94 \\
    & 1570 & 95.45 & 3.65 & 0.88 \\
\hline

0.6 & 1530 & 91.41 & 3.28 & 5.24 \\
    & 1550 & 93.98 & 4.07 & 1.89 \\
    & 1570 & 95.61 & 3.54 & 0.81 \\
\hline

0.7 & 1530 & 91.57 & 3.27 & 5.13 \\
    & 1550 & 94.19 & 3.89 & 1.87 \\
    & 1570 & 95.81 & 3.39 & 0.76 \\
\hline
\end{tabular}
\label{Table_S3}
\end{table}
\begin{table}[h]
\centering
\caption{Simulated modal fractions at the output monitor for $w = 5~\mu$m and $L = 200~\mu$m.}
\begin{tabular}{c c c c c}
\hline
Gap ($\mu$m) & Wavelength (nm) & TE0 (\%) & TE1 (\%) & TM0 (\%) \\
\hline

0.5 & 1530 & 58.94 & 10.08 & 30.26 \\
    & 1550 & 60.29 & 19.16 & 19.94 \\
    & 1570 & 62.87 & 28.88 & 7.84 \\
\hline

0.6 & 1530 & 57.57 & 11.09 & 30.65 \\
    & 1550 & 59.08 & 20.22 & 20.15 \\
    & 1570 & 61.84 & 29.88 & 7.95 \\
\hline

0.7 & 1530 & 56.29 & 11.78 & 31.18 \\
    & 1550 & 57.98 & 20.79 & 20.69 \\
    & 1570 & 60.89 & 30.46 & 8.36 \\
\hline
\end{tabular}
\label{Table_S4}
\end{table}
\begin{table}[h]
\centering
\caption{Simulated modal fractions at the output monitor for $w = 10~\mu$m and $L = 300~\mu$m.}
\begin{tabular}{c c c c c}
\hline
Gap ($\mu$m) & Wavelength (nm) & TE0 (\%) & TE1 (\%) & TM0 (\%) \\
\hline

0.5 & 1530 & 49.81 & 14.44 & 35.24 \\
    & 1550 & 54.95 & 24.83 & 19.73 \\
    & 1570 & 58.84 & 31.29 & 9.66 \\
\hline

0.6 & 1530 & 48.28 & 14.95 & 36.28 \\
    & 1550 & 53.67 & 25.45 & 20.46 \\
    & 1570 & 57.53 & 32.44 & 9.86 \\
\hline

0.7 & 1530 & 46.97 & 15.36 & 37.23 \\
    & 1550 & 52.47 & 25.88 & 21.31 \\
    & 1570 & 56.25 & 33.56 & 10.07 \\
\hline
\end{tabular}
\label{Table_S5}
\end{table}

Experimentally, mode purity is evaluated through resonance density over an extended wavelength range and multiple coupling gaps.
Figures 3d–g in the main text present representative spectra over a narrow wavelength range (1550--1553 nm) for the 0.6~$\mu$m gap. Here, we provide a more comprehensive analysis across coupling gaps and geometries.
For consistency across datasets, a fixed analysis window of 1540--1560 nm is applied to all spectra.
The wavelength window is converted to a frequency span using the speed of light c,
\begin{equation}
\Delta f = \frac{c}{\lambda_{\min}} - \frac{c}{\lambda_{\max}}.
\end{equation}
We extract a reference FSR (33.33 GHz) from a representative spectrum with the clearest resonance visibility. The expected number of fundamental-mode resonances within the analysis window is then
\begin{equation}
N_{\mathrm{expected}} = 
\frac{\Delta f}{\mathrm{FSR}_{\mathrm{ref}}}
= 74.88.
\end{equation}
The measured resonance count within the same wavelength window is denoted as $N_{\mathrm{measured}}$. 

We define the fundamental-mode fraction as
\begin{equation}
\eta = \frac{N_{\mathrm{expected}}}{N_{\mathrm{measured}}},
\end{equation}
which approximately reflects the proportion of resonances belonging to the fundamental mode family. Values slightly above unity can arise from discretization in the resonance counting.

\begin{table}[ht]
\centering
\caption{
Fundamental-mode fraction across coupling geometries and gaps}
\begin{tabular}{ccccc}
\toprule
$w$ ($\mu$m) & $L$ ($\mu$m) & Gap ($\mu$m) & 
$N_{\mathrm{measured}}$ & $\eta$ \\
\midrule

3  & 300 & 0.5 & 74 &  1.01\\
   &     & 0.6 &  74 &  1.01\\
   &     & 0.7 & 74 &  1.01\\
\midrule

5  & 300 & 0.5 & 75 &  1.00\\
   &     & 0.6 & 75 &  1.00\\
   &     & 0.7 &  74 &  1.01\\
\midrule

5  & 200 & 0.5 & 144  & 0.52 \\
   &     & 0.6 & 130 & 0.58 \\
   &     & 0.7 &  93 &  0.81 \\
\midrule

10 & 300 & 0.5 & 123 &  0.61\\
   &     & 0.6 & 134 & 0.56 \\
   &     & 0.7 & 110 & 0.68 \\
\bottomrule
\end{tabular}
\label{Table_S6}
\end{table}

We count resonances within a fixed wavelength window. Resonances are identified as local minima in the normalized transmission spectrum, under the threshold of $T_{\mathrm{norm}} < 0.95$. 
The measured number of resonances and corresponding $\eta$ are shown in Table \ref{Table_S6}. 
$\eta \approx 1$ across coupling conditions indicates preservation of a dominant TE$_0$ mode family, i.e., high mode purity.

\section{Evaluation of Coupling Ideality}

In the main text, we use the transmission model under ideal coupling as a qualitative framework 
to examine the device performance of different geometries 
($L = 200~\mu\mathrm{m}$ and $w = 5~\mu\mathrm{m}$ versus $w = 3~\mu\mathrm{m}$), 
and to motivate the choice of $w = 5~\mu\mathrm{m}$ as the primary coupling geometry. 
Here, we further clarify the physical meaning and scope of the coupling ideality concept. 

Coupling ideality was defined to account for the presence 
of parasitic coupling channels in realistic waveguide–resonator systems \cite{spillane2003_ideality}. 
In the ideal case, all coupled power is transferred into the desired fundamental mode, and the intrinsic loss rate is independent of the external coupling rate. 
In practice, additional parasitic channels may be present, 
such that a fraction of the coupled power is dissipated or coupled 
into unintended modes, resulting in extra effective loss that depends on the coupling conditions.

The coupling ideality factor $I$ is defined as the ratio between the power coupled 
into the desired fundamental mode and the total coupled power, including parasitic channels. 
Equivalently, in terms of decay rates,
\begin{equation}
I = \frac{\kappa_{e,0}}{\kappa_{e,0} + \kappa_p},
\end{equation}
where $\kappa_{e,0}$ denotes the intended external coupling rate and 
$\kappa_p$ represents parasitic coupling. An ideality factor of $I = 1$ corresponds to the ideal situation that all coupling occurs through the intended channel. 

There exist different conventions for incorporating the parasitic coupling rate $\kappa_p$ 
into the total linewidth $\kappa$. 
In some treatments, $\kappa_p$ is written explicitly, and $\kappa$ is expressed as
\begin{equation}
\kappa = \kappa_i + \kappa_{e,0} + \kappa_p,
\end{equation}
or equivalently in terms of $I$ as
\begin{equation}
\kappa = \kappa_i + \frac{\kappa_{e,0}}{I}.
\end{equation}
In other commonly used conventions, $\kappa_p$ is not written explicitly and is instead 
absorbed into an effective intrinsic loss $\kappa_i$.
In this representation,
parasitic coupling may appear as a coupling-dependent variation in the effective $\kappa_i$. 
Importantly, these different parameterizations do not alter the underlying physical behavior, 
but only affect how coupling non-idealities are expressed.

In this work, we adopt a simplified transmission-based evaluation of the coupling ideality. A full treatment of coupling behavior in practical devices can be substantially more complex.
We plot the expected relationship between the on-resonance transmission and the total dissipative linewidth under the ideal assumption that the intrinsic loss rate $\kappa_i$ remains constant and that no parasitic coupling channel $\kappa_p$ is present. Given $\kappa_e = \kappa - \kappa_i$, the on-resonance transmission follows
\begin{equation}
T_{\mathrm{ideal}}(\kappa)
=
\left(1 - \frac{2(\kappa - \kappa_i)}{\kappa}\right)^2.
\end{equation}
This representation provides a reference curve, such that experimental data can be visually compared ~\cite{pfeiffer2017_coupling}.
For Fig.~3h–i in the main text, the intrinsic loss rate $\kappa_i$ for the reference curve is estimated from devices with large coupling gaps and is set to be $2\pi \times 10.5 $ MHz for both cases. 

A limitation of transmission-based evaluation is that a single intrinsic loss rate $\kappa_i$ is used to generate the reference curve for resonances across different wavelengths and coupling gaps.
As discussed in the main text, the intrinsic loss rate follows a statistical distribution 
rather than being a single deterministic value. 
Consequently, individual data points are not expected to lie exactly on the ideal reference curve, but instead to approximately follow the overall trend.

To mitigate the influence of intrinsic loss variations, only resonances 
with total linewidth $\kappa$ below the median value are included 
in the comparison presented in Fig.~3h–i.
For the same coupling gap, the $w = 5~\mu\mathrm{m}$ geometry 
exhibits a smaller external coupling rate $\kappa_e$, 
consistent with its weaker effective coupling strength, 
leading to a correspondingly smaller total linewidth.
As a result, the accessible total linewidth range differs between the two geometries over the investigated coupling gaps (0.65–0.5~$\mu$m): the $w = 5~\mu\mathrm{m}$ devices span up to approximately 30~MHz, whereas the $w = 3~\mu\mathrm{m}$ devices extend toward 40~MHz. The minimum experimental coupling gap of 0.5~$\mu$m is set by our lithography limit.

Furthermore, the simplified coupling ideality model does not 
explicitly account for backscattering-induced mode splitting, characterized 
by $\kappa_c$. Variations in $\kappa_c$ across resonances can shift the 
measured on-resonance transmission away from the ideal expectation, 
introducing additional scatter in the data \cite{wang2024_coupling}. 
Overall, the coupling ideality model is employed here as a qualitative 
reference framework to evaluate device trends, rather than as a 
rigorous quantitative fitting model.

\section{Statistical Robustness and Bin Width Selection}

We examine the statistical robustness of the data presented in the main text with respect to the choice of histogram bin width.

Histograms provide a visual representation of the distribution of quantitative data, where the selection of bin width is a critical step. The bin width determines the interval range of each bin and defines how the entire data range is partitioned. While different choices of bin width may lead to slight variations in the extracted statistical parameters, these variations should remain limited under reasonable changes in bin width if the underlying distribution is robust.

A commonly used rule for selecting an appropriate histogram bin width is the Freedman--Diaconis rule \cite{freedman1981_histogram}. 
The bin width $h$ is estimated as
\begin{equation}
h = 2 \frac{\mathrm{IQR}}{n^{1/3}},
\end{equation}
where $\mathrm{IQR}$ denotes the interquartile range of the dataset and $n$ is the sample size.

For individual device datasets, the suggested bin width is approximately 1 MHz, whereas for the aggregated dataset of 10 devices, the recommended bin width decreases to approximately 0.47 MHz.
To maintain consistency across all analyses presented in this manuscript, we adopt a uniform bin width of 1 MHz for both single-device and multi-device histograms.
Although this choice results in a slightly larger bin width for the ensemble dataset, it effectively emphasizes the overall distribution trend while smoothing out minor fluctuations associated with finer binning.

\begin{figure*}[htbp]
    \centering
    \includegraphics[page=1,width=\textwidth]{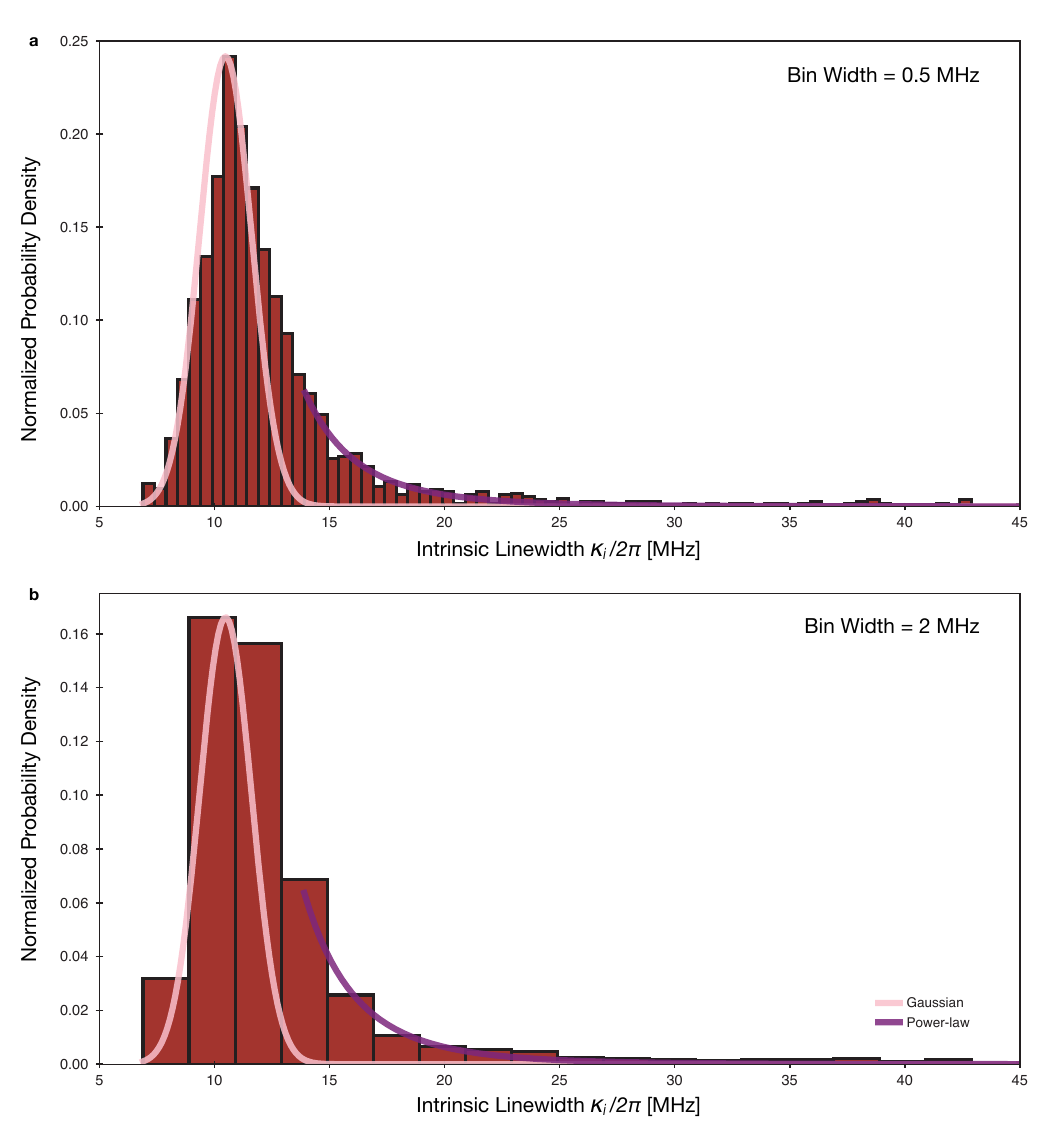}
    	\caption{\textbf{Robustness of histogram bin width selection.}
        Histograms of intrinsic linewidth $\kappa_i/2\pi$ constructed using different bin widths: \textbf{(a)} 0.5 MHz and \textbf{(b)} 2 MHz.
        Despite variations in visual smoothness, the Gaussian baseline (pink) and truncated power-law-like tail (purple) remain clearly distinguishable in both cases.}
    \label{Fig_S2}
\end{figure*}

We evaluated the robustness of the multi-device dataset (2,233 resonances) by varying the bin width to 0.5 MHz and 2 MHz. The corresponding histograms are shown in Fig. \ref{Fig_S2}.

While the visual appearance of the histogram changes slightly,
the Gaussian baseline and the power-law-like tail 
can be clearly distinguished.
The same fitting pipeline used in the main text is described in detail below, where we define 
$x = \kappa_i / 2\pi$:

1. A kernel density estimation (KDE) is applied to obtain a continuous estimate of the probability density of $x$ \cite{kde}.

2. A KDE-defined cutoff $h$ is identified from the dominant density peak. Data points satisfying $x \le h$ are selected as the baseline subset.

3. The baseline subset is modeled using a Gaussian distribution,
\begin{equation}
g(x) = \frac{1}{\sqrt{2\pi}\sigma}
\exp\!\left(-\frac{(x-\mu)^2}{2\sigma^2}\right),
\end{equation}
where $\mu$ and $\sigma$ are obtained via maximum-likelihood estimation.

4. The residual density beyond the Gaussian regime, defined for $x \ge \mu + 3\sigma$, is modeled using a truncated power-law,
\begin{equation}
r(x) = A x^{-\alpha} e^{-x/x_c},
\end{equation}
where $\alpha$ denotes the scaling exponent and $x_c$ represents the exponential cutoff scale.

Among the tested bin widths of 0.5 MHz, 1 MHz (used in the main text), and 2 MHz, the mode values corresponding to the highest histogram bins are 10.6 MHz, 10.4 MHz, and 9.9 MHz, respectively.
Although slight variations are observed, the differences remain within an acceptable range.

At the same time, the Gaussian fitting results remain unchanged, yielding $\mu = 10.48$ MHz and $\sigma = 1.13$ MHz for all three bin-width choices.
This consistency arises because the KDE-based cutoff and subsequent Gaussian fitting are independent of the histogram binning. Overall, these results confirm the robustness of our statistical analysis. 

\clearpage

\bibliography{references}
\bibliographystyle{sciencemag}

\end{document}